\documentclass[proceedings]{rmaa}

\usepackage{paralist}

\SetYear{2007}
\SetConfTitle{Proceedings of XII Latinamerican Regional IAU Meeting}

\title{Determining the Fractal Dimension\\
       of the Interstellar Medium}

\author{N\'estor S\'anchez,\altaffilmark{1}
        Emilio J. Alfaro,\altaffilmark{1} and
        Enrique P\'erez\altaffilmark{1}}

\altaffiltext{1}{Instituto de Astrof\'{\i}sica de Andaluc\'{\i}a,
                 CSIC, Apdo. 3004, E-18080, Granada, Spain
                 (nestor@iaa.es, emilio@iaa.es, eperez@iaa.es)}

\shortauthor{S\'anchez, Alfaro, \& P\'erez}
\shorttitle{Fractal Dimension of the ISM}
\listofauthors{N. S\'anchez, E. J. Alfaro, \& E. P\'erez}
\indexauthor{S\'anchez, N.}
\indexauthor{Alfaro, E. J.}
\indexauthor{P\'erez, E.}

\abstract{
The Interstellar Medium seems to have an underlying fractal structure,
which can be characterized through its fractal dimension ($D_f$).
However, several factors may affect the determination of $D_f$, such
as distortions due to projection, low image resolution, opacity of the
cloud, and low signal-to-noise ratio. Here we use both simulated clouds
and real molecular cloud maps to study these effects in order to estimate
$D_f$ in a reliable way. Our results indicate in a self-consistent way
that the fractal dimension of the Interstellar Medium is in the range
$2.6 \lesssim D_f \lesssim 2.8$, which is significantly higher than the
value $D_f \simeq 2.3$ usually assumed in the literature.}

\resumen{El Medio Interestelar parece tener una estructura fractal
subyacente, la cual se puede caracterizar a trav\'es de su dimensi\'on
fractal ($D_f$). Sin embargo, diversos factores pueden afectar la
determinaci\'on de $D_f$, tales como distorsiones causadas por la
proyecci\'on, baja resoluci\'on de la imagen, la opacidad de la
propia nube, y una baja raz\'on se\~nal/ruido. En este trabajo
utilizamos tanto nubes simuladas como mapas reales de nubes moleculares
para estudiar estos efectos y poder estimar de una manera fiable la
dimensi\'on fractal. Los resultados indican de una forma autoconsistente
que la dimensi\'on fractal del Medio Interestelar est\'a en el rango
$2.6 \lesssim D_f \lesssim 2.8$, lo cual es significativamente m\'as
alto que el valor $D_f \simeq 2.3$ que usualmente se asume en la
literatura.}

\addkeyword{ISM: clouds}
\addkeyword{ISM: individual (Ophiuchus, Orion, Perseus molecular cloud)}
\addkeyword{ISM: structure}

\begin{document}

\maketitle

\section{Introduction}

The fractal dimension ($D_f$) seems to be a simple but useful
tool for characterizing the degree of complexity of the
Interstellar Medium (ISM). Most of studies use the perimeter-area
method to calculate the fractal dimension of the boundaries of
projected molecular clouds ($D_{per}$). The observational evidence
can be summarized by saying that $D_{per}$ has a more or less
universal value around $\simeq 1.35$ (see \citealp{paper1} and
references therein). From this, it is usually assumed that the
real 3-dimensional fractal dimension is $D_f = D_{per} + 1 \simeq
2.35$ (e.g., \citealp{Bee92}), a value which moreover is consistent
with turbulent diffusion in an incompressible fluid \citep{Men90}.
In this contribution we show that this simple relation between $D_f$
and $D_{per}$ is not valid. We study carefully the effects that
different factors have on the determination of $D_f$ in order to
calculate in a highly reliable way its value for the ISM.

\section{Method}

Here we follow an empirical approach. We first simulate 3-dimensional
fractal clouds with a given well-defined fractal dimension $D_f$
(see details in \citealp{paper1,paper2}). Then we project the
clouds to generate 2-dimensional maps. When doing this, we consider
and analyze in detail the effects produced by the projection itself,
image resolution, cloud opacity, and low signal-to-noise ratio (S/N).
We take all these factors into consideration to develop algorithms
to estimate fractal dimensions in a precise and accurate way. We use
two different fractal estimators to corroborate the results and trends
obtained: the perimeter-area-based dimension ($D_{per}$) and the
mass-size dimension ($D_m$).

\section{Fractal dimension estimation}

Our main results can be summarized as follows.
\begin{asparaitem}
\item {\it Projection}. The projected perimeter dimension $D_{per}$
decreases as the 3-dimensional fractal dimension $D_f$ increases.
This can be seen in Figure~\ref{perimetro}
\begin{figure}[!t]
\includegraphics[width=\columnwidth]{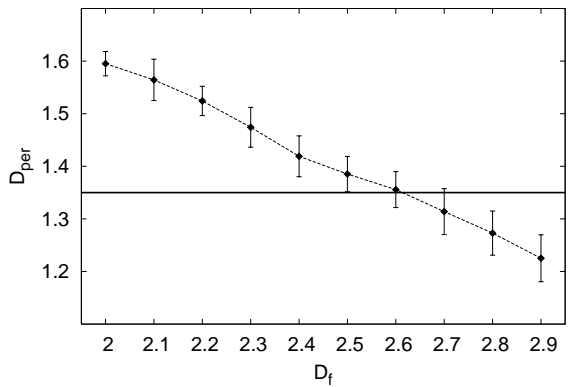}
\caption{Perimeter dimension $D_{per}$ as a function of the fractal
dimension $D_f$ calculated for 50 random clouds. Each data point is 
the average value and the bars are the standard deviations. The
horizontal solid line indicates the value $D_{per}=1.35$.}
\label{perimetro}
\end{figure}
for fractal clouds having dimensions in the range $2.0 \leq D_f \leq 2.9$.
The interesting point here is that the observed average value $D_{per}
\simeq 1.35$ is more consistent with $D_f \simeq 2.6$ than with $D_f
\simeq 2.3$.
\item {\it Resolution}. We have decreased the resolution by hand
and we have obtained that $D_{per}$ decreases as resolution decreases.
This is an expected result since as the pixel size is larger (worse
resolution) the perimeter becomes smoother because the fractal features
(the irregularities in the cloud contours) blend with each other. The
results shown in Figure~\ref{perimetro} correspond to ``high resolution"
images, i.e. cloud for which the size is $N_{pix} \geq 400$ pixels. In
this case we get $D_{per} = 1.36 \pm 0.03$ for the case $D_f=2.6$, but
if the resolution is decreased to $N_{pix} = 200$ then $D_{per} = 1.28
\pm 0.02$.
\item {\it Opacity.} We have generated cloud maps with different
values for the total optical depth $\tau$. We have obtained the
important result that opacity does not affect the estimation of
$D_{per}$. This is because even for optically thick clouds, for
which central and densest cores are difficult to detect, the
fractal properties of the cloud are contained in the shape of
the external contours. The situation is different for the mass
dimension $D_m$ because this estimator use information from all
the cloud structure (mass vs. radius). This method ($D_m$) fails
when $\tau \gtrsim 1$ but in any case it can be used without
problems for optically thin regions.
\item {\it Noise.} Very noisy maps can seriously affect the
estimation of both $D_{per}$ and $D_m$, artificially increasing
the structure irregularities and therefore decreasing the final
value of $D_f$. This effect can be minimized by smoothing the
map before calculating $D_f$. We have shown that this previous
step should be done in low-S/N maps only if the smoothing process
maximizes S/N.
\end{asparaitem}

\section{Application to the ISM}

\begin{figure}[!t]
\includegraphics[width=\columnwidth]{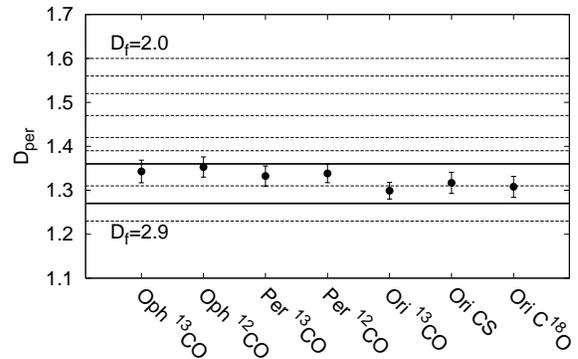}
\caption{Perimeter dimension $D_{per}$ obtained for each molecular
cloud map labeled in the horizontal axis. The horizontal lines
indicate values calculated for simulated clouds with $D_f$ values
from 2.0 to 2.9 in increments of 0.1. The interval $2.6 \leq D_f
\leq 2.8$ is emphasized with thick solid horizontal lines.}
\label{nubes}
\end{figure}

Taking all the above effects into account we have calculated the
fractal dimension of Ophiuchus, Perseus, and Orion molecular clouds
using emission maps in different lines. All the results are summarized
in Figure~\ref{nubes},
where we can see that the fractal dimension is always in the range
$2.6 \lesssim D_f \lesssim 2.8$. The mass dimension method
yielded very similar results \citep{paper3}.

These results are in perfect agreement with recent simulations of
compressively driven turbulence in the ISM \citep{Fed07}, and this
is an important point because, as pointed out by these authors, the ISM
is far to be an incompressible fluid. \citet{Fed07} obtained $D_f \simeq
2.6-2.7$ in their standard simulations and they showed that $D_f$ can
be as small as $\sim 2.4$ only for the extreme case of high turbulent
Mach numbers (of the order of 10). Another important point is that
it has been argued that a fractal ISM with $D_f \simeq 2.3$ could
account for some observed cloud properties (e.g., \citealp{Elm96}),
but we have also shown that $D_f \simeq 2.6$ is roughly consistent
with the average properties of the ISM, in particular the mass
and size distributions \citep{paper2}. In summary, it seems that
$D_f \simeq 2.7 \pm 0.1$ for the ISM. The relevance of this
relatively high fractal dimension value has to be analyzed,
mainly concerning the physical processes involved in the
structure of the ISM.

\end{document}